\documentclass[
reprint,
superscriptaddress,
amsmath,amssymb,
aps,
longbibliography
]{revtex4-2}

\usepackage{graphicx}% Include figure files
\usepackage{dcolumn}% Align table columns on decimal 
\usepackage{subfigure}
\usepackage{hyperref}
\usepackage{siunitx}
\usepackage{changes}
\usepackage{xcolor}

\newcommand{\ub}{{\bf u}}

\newcommand{\Ub}{{\bf U}}
\newcommand{\Xb}{{\bf X}}

\newcommand{\xb}{{\bf x}}
\newcommand{\rb}{{\bf r}}

\newcommand{\fb}{{\bf f}}
\newcommand{\Fb}{{\bf F}}

\newcommand{\turbTime}{\tau_\mathrm{turb}}
\newcommand{\St}{\mathit{St}}
\newcommand{\Rey}{\mathit{Re}}
\newcommand{\tumbTime}{\tau_\mathrm{tumb}}
\newcommand{\pdot}{\dot{\mathbf{p}}}
\newcommand{\pperp}{{\mbox{\tiny{$\perp$}}}}
\newcommand{\pparallel}{{\mbox{\tiny{$\parallel$}}}}

\begin{document}

\title{Fiber Tracking Velocimetry for two-point statistics of turbulence}

% authors
\author{Stefano Brizzolara}
\affiliation{Institute of Environmental Engineering, ETH Zurich, CH-8039 Z\"urich, Switzerland}
\affiliation{Swiss Federal Institute of Forest, Snow and Landscape Research WSL, 8903 Birmensdorf, Switzerland}

\author{Marco Edoardo Rosti}
\affiliation{Complex Fluids and Flows Unit, Okinawa Institute of Science and Technology Graduate University,\\
1919-1 Tancha, Onna-son, Okinawa 904-0495, Japan}

\author{Stefano Olivieri}%
\affiliation{Complex Fluids and Flows Unit, Okinawa Institute of Science and Technology Graduate University,\\
1919-1 Tancha, Onna-son, Okinawa 904-0495, Japan}
\affiliation{DICCA, University of Genova and INFN, Genova Section,
Via Montallegro 1, 16145, Genova, Italy}

\author{Luca Brandt}%
\affiliation{
Linne Flow Centre and SeRC (Swedish e-Science Research Centre),\\
Department of Engineering Mechanics, Royal Institute of Science and Technology, SE 100 44 Stockholm, Sweden}

\author{Markus Holzner}%
\affiliation{Swiss Federal Institute of Aquatic Science and Technology Eawag, 8600 D\"ubendorf, Switzerland}
\affiliation{Swiss Federal Institute of Forest, Snow and Landscape Research WSL, 8903 Birmensdorf, Switzerland}

\author{Andrea Mazzino}%
\affiliation{DICCA, University of Genova and INFN, Genova Section,
Via Montallegro 1, 16145, Genova, Italy}

\date{\today}

\begin{abstract}
We propose and validate a novel experimental technique to measure two-point statistics of turbulent flows. It consists in spreading rigid fibers in the flow and tracking their position and orientation in time and therefore been named ``Fiber Tracking Velocimetry'' (FTV). By choosing different fiber lengths, i.e. within the inertial or dissipative range of scales, the statistics of turbulence fluctuations at the selected lengthscale can be probed accurately by simply measuring the fiber velocity at its two ends, and projecting it along the transverse-to-the-fiber  direction. By means of fully-resolved direct numerical simulations and experiments, we show that these fiber-based transverse velocity increments are statistically equivalent to the (unperturbed) flow transverse velocity increments. Moreover, we show that the turbulent energy dissipation rate can be accurately measured exploiting sufficiently short fibers. The technique has been tested against standard Particle Tracking Velocimetry (PTV) of flow tracers with excellent agreement. Our technique overcomes the well-known problem of PTV to probe two-point statistics reliably because of the fast relative diffusion in turbulence that prevents the mutual distance between particles to remain constant at the lengthscale of interest. This problem, making it difficult to obtain converged statistics for a fixed separation distance, is even more dramatic for natural flows in open domains. A prominent example are oceanic currents, where drifters (i.e.~the tracer-particle counterpart used in field measurements) disperse quickly, but at the same time their number has to be limited to save costs. Inspired by our laboratory experiments, we propose pairs of connected drifters as a viable option to solve the issue.
\end{abstract}

\pacs{Valid PACS appear here}

\maketitle

\section{\label{Introduction}Introduction}
 
The physics of turbulent flows, which are ubiquitous in real-world applications, is a widely addressed, yet not fully understood, research topic, so that turbulence is often considered as the last unsolved problem of classical physics~\cite{frisch1995turbulence}. 
Examples where turbulence is important range from relatively small-scale problems (e.g., suspension dynamics~\cite{toschi2009lagrangian,voth2017review}, drag reduction~\cite{choi1994active,sirovich1997turbulent} or blood flow through heart valves~\cite{detullio2009direct,gulan2012experimental}) to large-scale geophysical phenomena in oceanography and meteorology~\cite{lacasce2008statistics,pouquet2013geophysical}. 
Overall, many experimental investigations rely on accurate measurements of turbulence, and consequently 
active research is devoted to the development of new methodologies able to access flow properties in a more detailed and convenient way.

In this paper, we propose a novel non-intrusive experimental technique based on tracking rigid fibers dispersed in a turbulent flow, which has been named ``Fiber Tracking Velocimetry'' (FTV). We believe that this method has great potential in the field of experimental turbulence; we expect it to be superior to the traditional methods based on tracer particles, in particular, when measuring the two-point statistics of turbulence. 
First, we expect this new method to overcome tracer-based methods when measuring inertial range scaling laws in situations where a high particle concentration is hardly reachable and/or maintainable in time. 
Second, we believe that it is more suitable to measure quantities related to spatial velocity gradients, such as the turbulent energy dissipation rate than the brute force approach of increasing the tracer concentration to reach the desired spatial resolution.
Indeed, the FTV method leads to a paradigmatic change, and overcomes the typical issues of tracer-based methods.

\subsection{Classical particle-based approaches}

Before introducing the FTV method, we briefly review the more currently used tracer-based techniques.
Standard measurement techniques are based on using \textit{tracers}, i.e. particles of typically micrometric size and density comparable to that of the fluid, so that their behavior is essentially the same of fluid particles. The two most popular methods are Particle Image Velocimetry (PIV)~\cite{adrian1991particle} and Particle Tracking Velocimetry (PTV)~\cite{maas1993particle}.

PIV is an optical technique that allows measurements of the instantaneous flow velocity field based on the displacement of tracer particles between two subsequent camera frames. The frames are split into interrogation areas and a displacement vector is obtained for each area using auto- or cross - correlation techniques. The displacement vectors are converted into velocity vectors using the time interval between two subsequent laser shots. Usually, the seeding particle concentration in PIV is such that it is possible to identify individual particles in an image, but not with enough certainty to track them between images. Exceptions are recent volumetric implementations of PIV (e.g. \citet{schroder2015advances}) that can also track individual particles.

Three dimensional particle tracking is typically conducted at lower seeding densities than PIV so that individual particles can be followed more easily over time~\cite{hoyer2005,krug2014,lawson2014}. Particles are usually tracked in 3D by using a stereoscopic camera arrangement~\cite{willneff2003spatio}. Based on the particle trajectories, 
one can evaluate the velocity vector by differentiating the particle coordinate vector in a Lagrangian setting, i.e. along the particle trajectories~\cite{luthi2005lagrangian}.

A well-known issue of tracer-based techniques is the evaluation of two-point statistical quantities, which are of special interest in turbulence. Indeed, in a turbulent flow the relative distance between two particles --- initially close enough --- grows in time as $t^{3/2}$, a phenomenon known as Richardson diffusion law~\cite{richardson1926atmospheric,lacasce2008statistics}. In other words, a significant problem when using particles to evaluate the statistics of turbulence is that two particles tend to separate from each other, thus making practical impossible to obtain converged statistics for a fixed separation distance. To tackle this issue and achieve reasonable statistics, measurements are usually conducted over a long time span, to have enough occurrences where couples of particles are found at a given distance~\cite{ouellette2009bulk,ni2011local,blum2011signatures,xi2013elastic}.
Such an approach becomes particularly burdensome when few particles can be seeded in the flow or when the flow domain is open (e.g. when conducting oceanographic measurements using drifters \cite{lacasce2008statistics}).

Another situation in which tracer-based methods manifest their intrinsic weakness is when measuring quantities related to the velocity gradient tensor, such as the turbulent dissipation rate. Currently, the only way to access such small-scale quantities is to track an extremely high number of particles. To give an idea of how challenging this is, one can consider that the flow field resolution is proportional to the inverse of the mean nearest neighbor distance, which depends on the particle concentration $n$ with the power-law $ \sim n^{1/3}$~\cite{hertz1909gegenseitigen}. In practice, this means that to increase ten times the flow field resolution, one needs to track one thousand times more particles, quickly leading to an unfeasible amount. For this reason, PTV-based estimation of the full velocity gradient tensor with a resolution close to the Kolmogorov scale has been mostly limited to moderate Reynolds number flows and small observation volumes~\cite{luthi2005lagrangian}. Finally, it is worth to mention that the number of degrees of freedom needed to properly describe a turbulent flow can be estimated as $Re^{9/4}$ \cite{frisch1995turbulence}: since the maximum number of traceable particles is technologically limited, the problem becomes particularly relevant when one needs to measure small-scale quantities in highly inertial flows.

\subsection{Fiber-based approach}

In light of the weaknesses associated with tracer-based methods, here we propose an alternative strategy based on using approximately one-dimensional fiber-like objects instead of tracer particles. The reason for considering fibers instead of particles is rather simple and relates to the fact that the distance between the two ends of the fiber is constant. As it will be shown, this is the essential feature by virtue of which it is possible to investigate the behavior of turbulent eddies of a selected size. Conversely, this is not easy to achieve when considering a pair of fluid tracers due to the above mentioned Richardson dispersion.

While a significant research effort has focused on understanding the dynamics of anisotropic particles of size smaller than the Kolmogorov lengthscale (i.e., the smallest scale where viscous dissipation is predominant)~\cite{parsa2012,ni_kramel_ouellette_voth_2015,sabban_cohen_van-hout_2017,voth2017review}, the dynamical behavior of fibers of finite length (i.e., within the inertial range) started to be investigated only recently, both experimentally and numerically~\cite{parsa2014,bounoua2018tumbling,rosti2018flexible,rosti2019flowing,pujara_voth_variano_2019,olivieri2020dispersed}. 

The main result in literature of relevance here, is that finite-length fibers (i.e., fibers whose length falls within the inertial range) rotate (rigid fibers)~\cite{bounoua2018tumbling} or deform (flexible fibers)~\cite{rosti2018flexible} at the same frequency of the turbulent eddies of comparable size. Previous studies mostly focused on the so-called tumbling rate whose scaling showed reasonable agreement with predictions in the inertial range \cite{parsa2014,bounoua2018tumbling}. However, the more general question as to how fibers can be used as a proxy of other dynamical properties in the inertial and viscous range remained open. As we will demonstrate, quantitative information of two-point statistics of turbulence can be obtained by tracking the fiber motion in time and looking at the velocity difference between the fiber ends. While it is expected that velocity differences at the fiber ends coincide with flow velocity differences measured at the same two ends because of the no-slip condition, the nontrivial result we will reveal here is that the fibers can measure the \textit{undisturbed} flow velocity differences. This unravels the potential of having non-intrusive measures of two-point properties of turbulence. Fibers therefore appear as ideal candidates for investigating the inertial range (IR) of turbulence, as well as the viscous scales that are usually difficult to access with tracer-based methods. In this paper, we prove that the motion of rigid fibers is strongly linked with the turbulent eddy motion, both in the inertial and viscous range and use this to conceive a novel experimental technique.

Our technique takes advantage of the interconnection between fiber motion and fluid flow across scales, providing a novel way to probe inertial velocity structure functions and the turbulent dissipation rate. Such an approach was not suggested by previous studies that mostly focused on small point-like fibers in turbulence.

\subsection{Outline of the work}
In this work, we use rigid fibers to perform measurements of two-point turbulence statistics.
From a practical viewpoint the choice of rigid, instead of flexible, fibers comes from the easier fabrication and control of their physical properties. Moreover, when fibers are rigid the data acquisition and processing is strongly simplified.
In the following, we first introduce the theoretical background providing the basis for the validity of the proposed method (Sec.~\ref{sec:arguments}). Our main arguments are validated by means of direct numerical simulations of a rigid fiber immersed in a homogeneous isotropic turbulent flow (Sec.~\ref{sec:DNS}). The foundations to realize the experimental approach of a ``Fiber Tracking Velocimetry" technique are introduced in Sec.~\ref{sec:experimental_method}. In Sec.~\ref{sec:results} we compare the results obtained through Fiber Tracking Velocimetry and a benchmark measurement of 3D particle tracking. We conclude the manuscript with a summary of the main findings and outline future developments (Sec. \ref{conclusion}).

\section{Relevant statistical observables}
\label{sec:arguments}

\begin{figure}[t]
\includegraphics[width = \columnwidth]{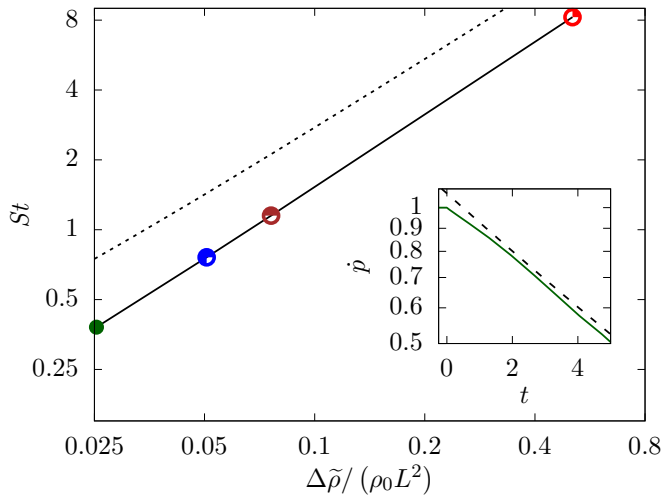}
\caption{\label{Stokes} Stokes number as a function of the linear density difference evaluated from the numerical simulations. Symbols and colors represent fibers with different values of $\Delta\widetilde{\rho}$. The dashed line reports the prediction obtained from Eq. \eqref{Stk_eq}. The inset shows the time history of the fiber angular velocity in one of our numerical test (solid line) along with the exponential fitting curve (dashed line).}
\end{figure}

In this section, we start by providing the essential concepts and defining the main quantities involved in this work.

We consider a rigid fiber with length $c$ and diameter $d$, such that its aspect ratio (or slenderness) is $c/d \gg 1$. Neglecting gravitational effects (i.e. the Froude number is assumed to be negligible), the main dimensionless parameter governing the dynamical behavior of the fiber is the rotational Stokes number
\begin{equation}
\label{eq:St_direct}
St = \frac{\tau_\mathrm{r}}{\tau_\mathrm{f}},
\end{equation}
which is defined as the ratio between the fiber response (or relaxation) time  $\tau_\mathrm{r}$ and the characteristic timescale of the flow $\tau_\mathrm{f}$, and therefore represents a measure for the inertia of the dispersed object.
 
In the simulations, the value of $\St$ can either be measured or estimated.
We evaluate the fluid timescale as the time a fluid particle takes to perform a complete circular trajectory of diameter $c$ flowing with a velocity $\delta u_\pperp / 2$, thus yielding  $\tau_\mathrm{f} = 2 \pi c / \delta u_\pperp$.
On the other hand, the fiber response time is obtained by measuring the rotation decay rate of a fiber initially rotating at a constant angular speed in a quiescent fluid.
% Using the same physical properties of the reported DNS, 
An example of the resulting time history is shown in the inset of Fig.~\ref{Stokes}, on which an exponential fit is applied to obtain the response time. The test is performed for four different fibers of length $c = 0.16 L$ and four different values of the linear density difference $\Delta \widetilde{\rho}$. The resulting Stokes number is plotted as a function of $\Delta \widetilde{\rho}$ in the main figure, showing a linear proportion that is consistent with the prediction based on the slender body theory~\cite{cavaiola_olivieri_mazzino_2020}.

In the experiments, a similar evaluation of the Stokes time is not feasible; hence we choose to estimate the Stokes number using the semi-empirical relation proposed by~\citet{bounoua2018tumbling} for rigid fibers in turbulent flows which reads as:
\begin{equation}
\mathit{St} = \frac{1}{48} \frac{\rho_\mathrm{s}}{\rho_\mathrm{f}} \left( \frac{d}{\eta}\right)^\frac{4}{3} \, \left( \frac{d}{c}\right)^\frac{2}{3}   \left[ 1 + \frac{3}{4}  \left( \frac{d}{c} \right)^{2}  \right],
\label{Stk_eq}
\end{equation}
where $\eta$ is the Kolmogorov lengthscale. To examine the reliability of this relationship, we compare the Stokes number directly obtained via DNSs with the prediction of Eq. (\ref{Stk_eq}). Fig. \ref{Stokes} shows that, despite the same scaling is obtained, Eq. (\ref{Stk_eq}) overestimates the Stokes number of a factor $\sim 1.5$. Note that we are not interested in the exact value of the Stokes number but we employ this relation only to verify that in our experiments $\St \ll 1$. In this sense, Eq. (\ref{Stk_eq}) provides a precautionary estimation of $St$.

When tracking the fibers during their motion, we will focus on their rotation rate, which can be decomposed into spinning and tumbling degrees of freedom~\cite{voth2017review}. The latter, in particular, is proportional to the rate of change of the fiber orientation vector $\pdot$. Hence, we can define the tumbling time as
\begin{equation}
\tumbTime = \langle \pdot \cdot \pdot \rangle^{-1/2},
\label{eq:tumbTime}
\end{equation}
representing a measure of the characteristic timescale at which the rigid fiber is moving in the turbulent flow. 

For the fluid flow, let us recall some well-known results from Kolmogorov theory (1941, denoted hereafter as K41), assuming the case of homogeneous and isotropic turbulence and initially focusing on the inertial range of scales~\cite{frisch1995turbulence,pope2000turbulent}.
The first is the scaling of the characteristic timescale associated with turbulent eddies of size $r$ (the so-called eddy-turnover time) that can be expressed as
\begin{equation}
\turbTime(r) \sim \epsilon^{-1/3} \, r^{2/3},
\label{eq:turbTimeScaling}
\end{equation}
where $\epsilon$ is the turbulent energy dissipation rate.
Eq.~\eqref{eq:turbTimeScaling} follows from the celebrated $4/5$ law, one of the few exact results available in turbulence, which relates the third-order longitudinal structure function $S_3^\pparallel(r) = \langle \delta u_\pparallel ^3 \rangle$, where $\langle \cdot \rangle$ denotes ensemble average of the longitudinal velocity increment $\delta u_\pparallel (r)= [\ub(\xb+\rb,t) - \ub(\xb,t)] \cdot \hat{\rb}$, to the separation distance:
\begin{equation}
S_3^\pparallel = - \frac{4}{5} \, \epsilon \, r,
\label{eq:S3parallel}
\end{equation}
where $r = |\rb|$ and $\hat{\rb} = \rb / r$ is the unit vector along the separation.

Unfortunately, when dealing with rigid fibers as in the present work such longitudinal projection evaluated using the fiber velocities is always zero. On the other hand, the possibility of measuring the longitudinal velocity increments using flexible fibers has been explored numerically in Refs.~\cite{rosti2018flexible,rosti2019flowing}. In the rigid case, nevertheless, fibers remain a good candidate for probing the flow if one focuses on transverse, instead of longitudinal, velocity differences, i.e. considering the projection of the velocity difference along an orthogonal direction: 
$\delta u_\pperp = [\ub(\xb+\rb,t) - \ub(\xb,t)] \cdot \hat{\rb}_\pperp$, with $\hat{\rb}_\pperp$ being a unit vector normal to $\rb$ \citep{cavaiola_olivieri_mazzino_2020}. As it will be shown later, this is the case if the fiber rotational Stokes number is sufficiently small.

Overall, we argue that, if the resulting tumbling time is comparable with the eddy-turnover time of turbulence at the same scale, i.e. $\tumbTime \approx \turbTime (r=c)$, then the fiber behaves as a proxy of turbulence. If this is the case, fiber tracking can be used to measure the transverse velocity difference and relevant statistical two-point quantities such as its probability density function (PDF), as well as the transverse structure functions $S_p^\pperp = \langle \delta u_\pperp ^p \rangle$ of any order $p$.
Note that, in homogeneous isotropic turbulence, the $4/5$th Kolmogorov law, yields a zero transverse third-order structure function $S_3^\pperp$. It then follows that the PDF of the transverse increments is symmetric (i.e. its skewness is zero, contrarily to the negative skewness of the PDF of the longitudinal velocity increments). To avoid this trivial result, in the following we will consider $\widetilde{S}_3^\pperp$ based on the absolute value of  $\delta u_\pperp$.
Note also that, although there is no equivalent analytical result to the longitudinal velocity increment, for the second-order transverse structure function phenomenological arguments lead to~\cite{pope2000turbulent}
\begin{equation}
S_2^\pperp = \frac{4}{3} \, \mathcal{C}_2 \, \epsilon^{2/3} \, r^{2/3},
\label{eq:S2_tra}
\end{equation}
where $\mathcal{C}_2$ is the so-called Kolmogorov constant. Such quantity generally depends on the particular flow configuration; in the case of homogeneous isotropic turbulence, we have $\mathcal{C}_2 = 2.1 \pm 0.5$~\cite{noullez1997transverse}. Furthermore, using $S_2^\pperp$ we can refine the expression of the eddy turnover time as
\begin{equation}
\turbTime(r) = \frac{r}{\sqrt{\frac{15}{2} S_2^\pperp}},
\label{eq:turbTime}
\end{equation}
from which it is easy to verify that $\lim_{r \to 0} \turbTime(r) = \tau_\eta = (\nu/\epsilon)^{1/2}$~\cite{pope2000turbulent},  $\eta$ being the Kolmogorov lengthscale and $\tau_\eta$ the Kolmogorov timescale.

Finally, we consider the turbulent energy dissipation rate $\epsilon=2\nu\,\langle \frac{\partial u_i}{\partial x_j} \frac{\partial u_i}{\partial x_j} \rangle$. For homogeneous isotropic turbulence, the latter reduces to
\begin{equation}
\epsilon = \frac{15}{2} \nu \, \bigg \langle \left( \frac{\partial u_1}{\partial x_2} \right)^2 \bigg \rangle,
\label{eq:epsilon}
\end{equation}
where $\nu$ is the kinematic viscosity of the fluid and where the spatial derivative of the velocity component $u_1$ is performed along the orthogonal direction $x_2$. While tracking the fiber, the latter expression can be transposed into the Lagrangian framework to evaluate the dissipation rate by approximating the spatial derivative appearing in Eq. \eqref{eq:epsilon} with the ratio between the normal velocity difference measured at the fiber ends and the fiber length, i.e. 
\begin{equation}
\epsilon \approx \frac{15}{2} \nu \, \bigg \langle \left( \frac{\delta u_\pperp}{c} \right)^2 \bigg \rangle. %  \quad \textrm{with} \,\, i \ne j,
\label{eq:epsilonFib}
\end{equation}
Note that, this approximation holds as long as $c$ is small enough; in order to provide a reliable measure of $\epsilon$ the fiber length should be comparable to $\eta$.

\section{Numerical evidence}
\label{sec:DNS}
Before introducing the experimental method, we present results from fully-resolved direct numerical simulations (DNS) where a single rigid fiber is freely moving in homogeneous isotropic turbulence (details on the numerical method are given separately in Appendix~\ref{app:DNS}). Simulations have been performed for different values of the fiber length $c$ and of the Stokes number $\St$ (see Fig.~\ref{Stokes}). For all the simulations, the Reynolds number based on the Taylor microscale is equal to $\Rey_\lambda \approx 92$ and the turbulent dissipation rate (normalized with the root-mean-square velocity and the box size $L$) is $\epsilon \approx 2.5$.

\begin{figure}[t]
\includegraphics[width = \columnwidth]{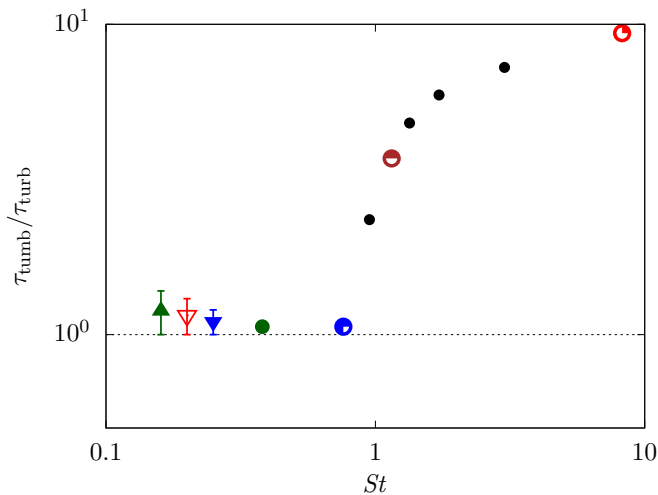}
\caption{\label{fig:tumb_DNS} Ratio between the fiber tumbling time (evaluated from Eq. \eqref{eq:tumbTime}) and the eddy turnover time at the fiber lengthscale (evaluated from Eq. \eqref{eq:turbTime}) as a function of the Stokes number. Colored bullets refer to the simulation data, while colored triangles are the experimental data. Coloring and symbols correspond to the values of $\St$ used also in the following figures.}
\end{figure}

Fig.~\ref{fig:tumb_DNS} reports the measured fiber tumbling time normalized with the eddy turnover time (evaluated at the fiber lengthscale $r=c$), as a function of the Stokes number. We notice that, for sufficiently small $\St$, the curve shows a horizontal plateau and the ratio becomes nearly unitary, indicating that the fiber is rotating at the same frequency of the turbulent eddies.
On the other hand, for larger $\St$ the tumbling time is found to be larger than the turbulent one, i.e. inertia causes the fiber response to be substantially delayed with respect to the flow variations. These findings are consistent with those recently reported by~\cite{parsa2014,bounoua2018tumbling,pujara_voth_variano_2019,kuperman2019inertial, bordoloi2020lagrangian} (note that the tumbling time plotted in Fig.~\ref{fig:tumb_DNS} is proportional to the inverse of the square root of the tumbling rate reported by \cite{parsa2014,bounoua2018tumbling,pujara_voth_variano_2019}). 
For larger $\mathit{St}$, it is difficult to distinguish a clear scaling, although a qualitative agreement is present.
Note also that the typical modeling approach makes use of several simplifications such as neglecting the effect of fiber inertia and the back-reaction of the fibers on the flow (one-way coupling). In our case, instead, the two phases are fully coupled and the effect of fiber inertia is accounted for.

\begin{figure}[t]
\includegraphics[width = \columnwidth]{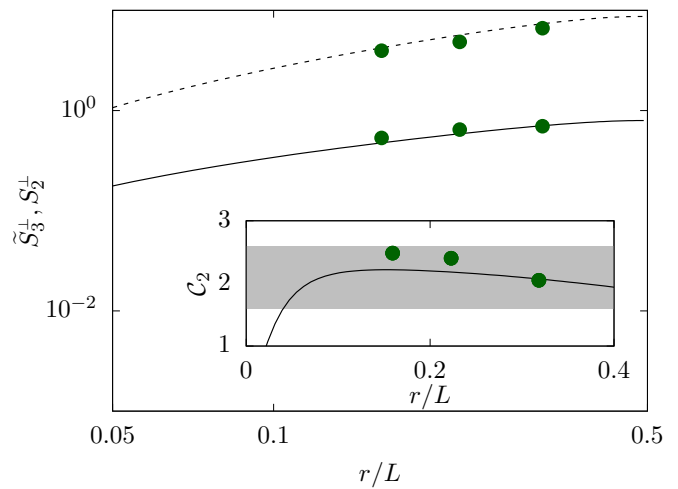}
\caption{\label{fig:S23_DNS} Second-order and third-order transverse velocity structure functions from DNS (note that $\widetilde{S}^\pperp_3$ is computed using the absolute value of the velocity difference). Solid and dashed lines indicate the standard Eulerian measurements for $S_2^\pperp$ and $\widetilde{S}_3^\pperp$, respectively, while symbols denote those obtained from Lagrangian tracking of fibers with $\St = 0.4$ and different lengths $c/L = 0.16$, $0.22$ and $0.32$. All quantities are made dimensionless with the box size $L$ and the velocity root-mean-square; $\widetilde{S}_3^\pperp$ is multiplied by $10$ to enhance the visibility. The inset shows the Kolmogorov constant $\mathcal{C}_2$ computed according to Eq. \eqref{eq:S2_tra} using both methods; the grey area denotes the range of values determined experimentally by~\citet{noullez1997transverse}.
} 
\end{figure}

\begin{figure}[t]
\includegraphics[width = \columnwidth]{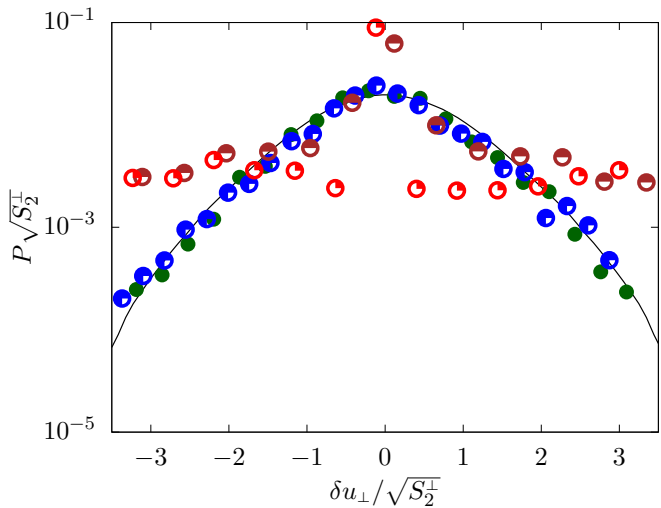}
\caption{\label{pdf_tra_dns} Probability density function (PDF) of the transverse velocity difference for fiber length (or separation distance) $c/L = 0.16$; the bullets correspond to the four different fiber linear densities reported in Fig.~\ref{Stokes}, while the black solid line is the Eulerian PDF.} 
\end{figure}

In order to exploit the fiber as a proxy of the flow, we necessarily need to have a sufficiently small Stokes number, so that $\tumbTime \approx \turbTime$. Here, we choose and retain $\St = 0.4$. Being this condition satisfied, we compute the second-order and third-order transverse velocity structure functions using the velocity difference between the two fiber ends, and compare such Lagrangian measurement with the more traditional one based on using the fluid velocity on the Eulerian grid. Results are shown in Fig.~\ref{fig:S23_DNS}, where three different fiber lengths are used for accessing different separation distances. A good agreement between the Eulerian measurements and those obtained from the fiber tracking is evident.

Next, recalling Eq. \eqref{eq:S2_tra}, we can exploit our data also to extract the value of the Kolmogorov constant, i.e. $\mathcal{C}_2 = S_2^\pperp / (4/3 \, \epsilon^{2/3} \, r^{2/3})$, which is reported in the inset of Fig.~\ref{fig:S23_DNS} as a function of the separation distance.
The data show that the fiber-based measurement is comparable within the uncertainty bounds with the literature results~\cite{noullez1997transverse}.

In Fig.~\ref{pdf_tra_dns} we show the PDF of the transverse velocity increment computed using the velocity difference between the fiber ends for different Stokes numbers, along with the same quantity evaluated in the Eulerian way. Results show that only those fibers with a sufficiently small $\St$ are able to reproduce the same PDF of the undisturbed carrier flow, while as $\St$ grows the agreement between the Lagrangian and Eulerian measure gets worse. 

\begin{figure}[t]
\includegraphics[width = \columnwidth]{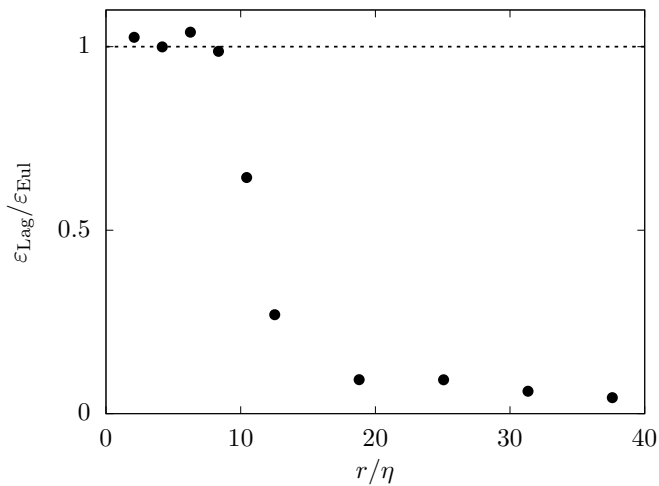}
\caption{\label{fig:eps-num} Turbulent dissipation rate (plotted as the ratio between the Lagrangian and Eulerian measurement) as a function of the separation distance $r$ normalized with the Kolmogorov lengthscale $\eta$.} 
\end{figure}

Finally, using Eq. \eqref{eq:epsilon} we exploit the fiber to measure the turbulent dissipation rate. Fig.~\ref{fig:eps-num} shows the ratio between the Lagrangian measurement $\epsilon_\mathrm{Lag}$ and the Eulerian one $\epsilon_\mathrm{Eul}$ as a function of the separation distance normalized with the Kolmogorov lengthscale $r/\eta$.
As clearly shown in the figure, the Lagrangian measurement provides a good estimate of the correct value of $\epsilon$ only for sufficiently short fibers with $c \lesssim 10 \eta$ while it gives a strong underestimation for longer fibers.

It is worth noticing that the two fiber ends do not behave as fluid tracers (also when the condition of small Stokes or Taylor, number is attained), due to the fiber inextensibility condition. Nevertheless, the effect of fiber inextensibility is excluded by projecting the end-to-end velocity difference along a normal direction, thus obtaining the so-called transverse velocity difference. Moreover, the flow modification due to the presence of the fiber turns out to be negligible provided that the Stokes number is sufficiently small~\cite{cavaiola_olivieri_mazzino_2020}.

\section{\label{sec:experimental_method}Experimental method}

In light of the evidence provided by the numerical study, a laboratory experiment has been designed and realized to prove the actual feasibility of measuring turbulence properties by the Lagrangian tracking of rigid fibers. The experiment consists in generating a three dimensional controllable turbulent flow within a water tank and suspending rigid fibers with a small Stokes number. By tracking fibers' edge positions in time, the turbulence observables defined in Sec.~\ref{sec:arguments} are evaluated, and compared with benchmark PTV results of flow tracers.

\begin{figure}[t]
\includegraphics[width = \columnwidth]{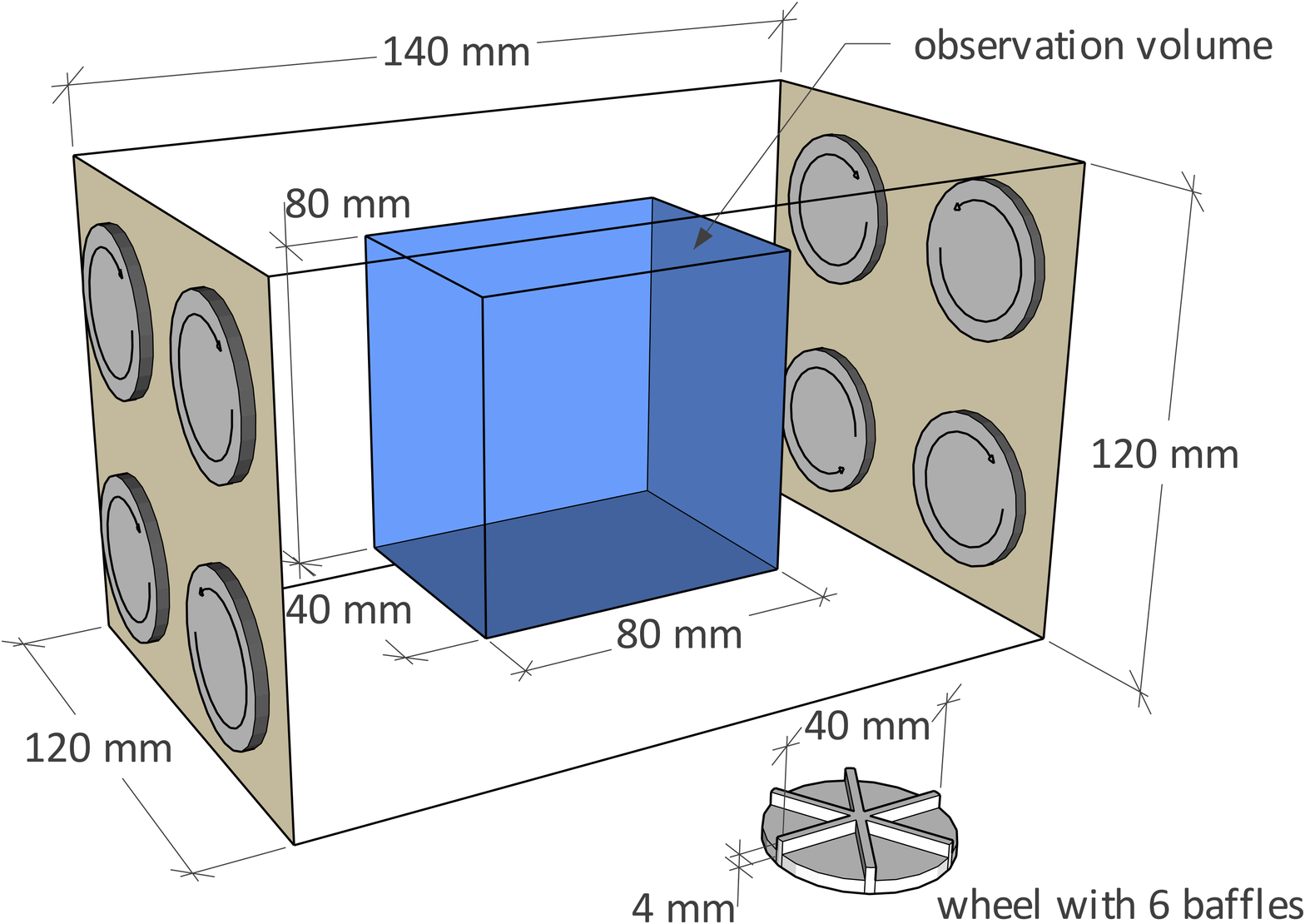}
\caption{\label{fig:setup} Schematic representation of the experimental setup: overview of the aquarium with turbulence generator and observation volume and close-up on one of the actuated wheels.}
\end{figure}

%\subsection{\deleted{Experimental set-up}}
In order to validate the Fiber Tracking Velocimetry (FTV) technique in a controlled environment, an approximately homogeneous and isotropic turbulence is generated. To this purpose, the same apparatus employed in~\cite{liberzon2005turbulence} and \cite{hoyer2005} is used to force the fluid motion. This consists in an aquarium of $120 \times 120 \times 140 \,\mathrm{mm}^3$ filled with water, and equipped with a turbulence generator. The turbulence is sustained by two sets of four wheels covered by artificial rough elements. The wheels counter-rotate according to the scheme shown in Fig.~\ref{fig:setup}. The rotation is driven by a closed loop controlled servo-motor installed on top of the aquarium. The rough wheels can be replaced by smooth wheels to generate flow field with a lower turbulence intensity \cite{michalec2017zooplankton}. The observation volume, which is approximately $80 \times 80 \times 60 \, \mathrm{mm}^3$, is located midway between the wheels. A coherent laser beam is used to illuminate the aforementioned field of view from below. Four synchronized high-speed cameras are used to record a stereoscopic view of the observation volume; the data are stored in real time on two fast-writable hard disks and then transferred to traditional supports for further analysis.

A complete characterization of the flow field is obtained using the more traditional particle tracking velocimetry (PTV) technique. For this purpose, the flow is seeded with white reflective, non-fluorescent and neutrally-buoyant particles that are tracked using the open source OpenPTV software (\url{https://www.openptv.net})\cite{maas1993particle}. Particles have a mean diameter of approximately $40 \, \mathrm{\mu m}$, and $10$ particles per $\mathrm{cm}^3$ are suspended in the flow. The result of this operation is a series of particle trajectories, from which the Lagrangian velocity and acceleration can be obtained, and finally mapped into an Eulerian grid.

In order to perform fiber tracking velocimetry, $50$ fibers are released into the flow such that their maximum local concentration is approximately $7.5\times10^{-2}$ fibers per $\mathrm{cm}^3$. This results in an extremely dilute configuration ensuring that the flow statistics are not modified by the presence of the dispersed phase. Two different types of fibers are fabricated: i) polydimethylsiloxane (PDMS) fibers tagged with a fluorescent dye at the edges, and ii) Nylon fully dyed fibers. In both cases the fibers edge positions and velocities are determined using the OpenPTV software; when using fully dyed Nylon fibers, a specific image segmentation procedure is applied to detect the fiber edges in the image space. For a detailed explanation of the fabrication and tracking procedure, the reader is referred to Appendix \ref{sec:Fiber fabrication} and \ref{sec:Fiber tracking technique} respectively.

\section{Experimental verification}
\label{sec:results}
In the following, a complete comparison between the PTV and FTV technique is carried out for the observables introduced in Sec.~\ref{sec:arguments}. The transverse velocity difference probability density function, the second and third order moment scaling are used as a proxy to check the FTV reliability within the inertial range of scales. We use the average turbulent dissipation rate to verify the reliability of the method in capturing the flow properties at the Kolmogorov scale. It is worthy to underlying that the experiment had been carried out in a set-up that allows both the PTV and FTV techniques, in order to test the reliability of the latter.

\begin{figure}[t]
\centering
\includegraphics[width = \columnwidth]{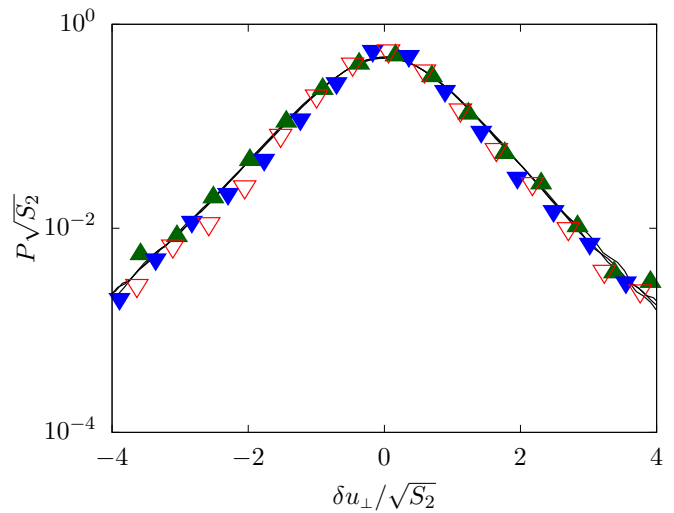}
\caption{PDF of the transverse velocity difference evaluated with particle tracking (solid lines) and Fiber Tracking Velocimetry (symbols) for $c/\mathcal{L} = 0.71$ (dark green), $c/\mathcal{L} = 0.45$ (blue) and $c/\mathcal{L} = 0.57$ (red), where $\mathcal{L}$ is the integral length scale. The distributions are normalized with the variance of the velocity differences at the corresponding separation, evaluated through the standard 3D-PTV technique. Turbulence is generated by rough wheels rotating at $400\,\mathrm{rpm}$. The corresponding Reynolds number based on the Taylor microscale $Re_{\lambda} = 146$, while the one based on the integral length scale $Re_{\mathcal{L}} = 1410$. Other relevant parameters are summarized in \citep{turbPropertiesNote}.}
\label{fig:pdf_tra_exp}
\end{figure}

\begin{figure}[t]
\includegraphics[width = \columnwidth]{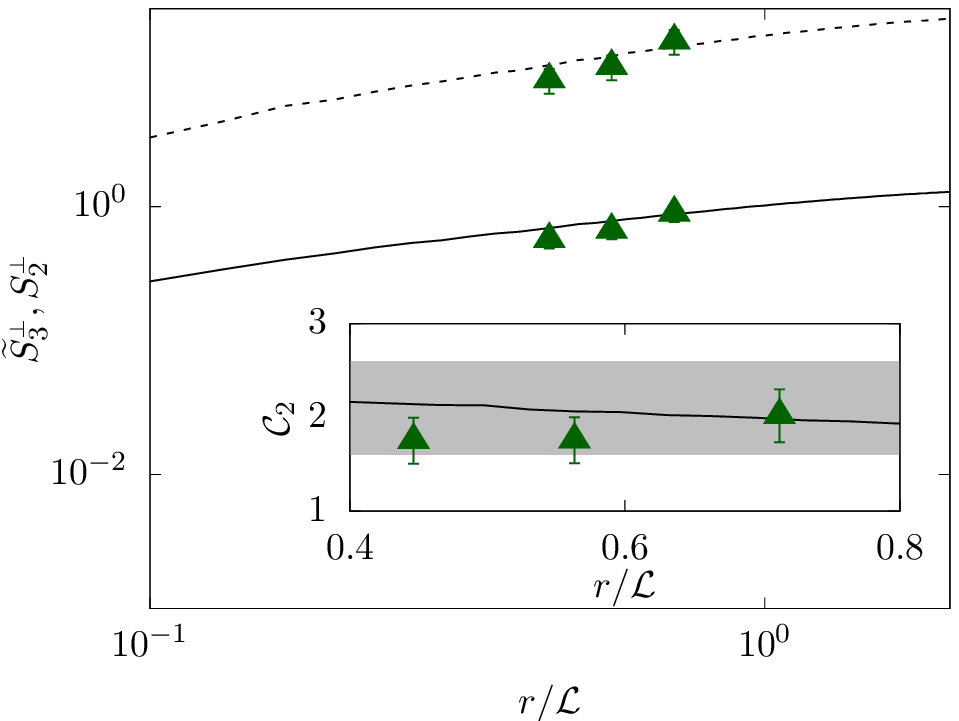}
\caption{\label{fig:S23_exp} Second-order and third-order transverse velocity structure functions from experiments (note that $\widetilde{S}^\pperp_3$ is computed using the absolute value of the velocity difference). Solid and dashed lines indicate the standard particle tracking measurement for $S_2^\pperp$ and $\widetilde{S}_3^\pperp$ respectively, while symbols denote those obtained from Lagrangian tracking of the fibers. The bars represents the error related to the velocity measurement. All quantities are made dimensionless with the integral scale $\mathcal{L}$ and the velocity root-mean-square;  $\widetilde{S}_3^\pperp$ is multiplied by $10$ to enhance the visualization. The inset shows the Kolmogorov constant $\mathcal{C}_2$ computed according to Eq. \eqref{eq:S2_tra} using both methods; the grey area denotes the range of values determined experimentally by~\citet{noullez1997transverse}. Turbulence is generated by rough wheels rotating at $400\,\mathrm{rpm}$.}
\end{figure}

\paragraph*{Within the inertial range.}
Fig.~\ref{fig:pdf_tra_exp} shows the PDF of the transverse velocity differences evaluated both with particles and PDMS fibers. Here, in order to avoid non-convergence problem in the density estimation, some extreme events are discarded removing data characterized by a low probability of $10^{-4}$. As seen from these three PDFs, the FTV provides results similar to the PTV. In Fig.~\ref{fig:S23_exp}, the second and third order structure function of the transverse velocity differences $S_2^{\pperp}$ and $\widetilde{S}_3^\pperp$ are shown: the results from the FTV (dark-green points) show a scaling behavior that is comparable with the PTV data. The inset of Fig.~\ref{fig:S23_exp} shows the Kolmogorov constant $\mathcal{C}_2$ of Eq. (\ref{eq:S2_tra}): the FTV measure (dark-green points) is contained within the range of values determined by \citet{noullez1997transverse}.

\begin{figure}[t]
\includegraphics[width = \columnwidth]{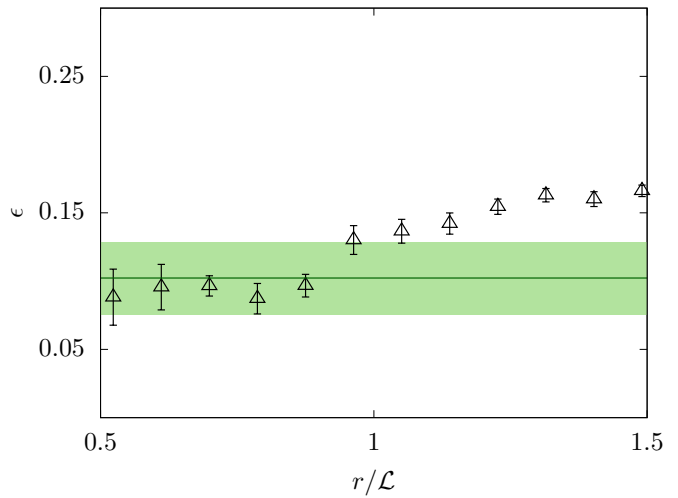}
\caption{\label{fig:epsilonfib} Comparison between the turbulence dissipation rate evaluated with FTV and PTV. The green line represents the value obtained by Eq. \eqref{eq:epsilon} from the Nylon fibers, and the green shadow is the relative tracking error. The black triangles are evaluated through the PTV data as $\epsilon = - S_3^\pparallel / \left( \frac{4}{5} \,  \, r \right) $, relying on the turbulence isotropy. The bars represent the error related to statistical convergence. All quantities are made dimensionless with the integral scale $\mathcal{L}$ and the root mean-square velocity. The turbulence is generated by smooth wheels rotating at $150\,\mathrm{rpm}$ (a complete characterization of the turbulence properties generated by smooth wheels can be found in \citep{michalec2017zooplankton}).}
\end{figure}

\paragraph*{At the Kolmogorov scale.}
The average turbulent dissipation rate $\epsilon$ is estimated using Nylon slender fibers $3\,\mathrm{mm}$ long (shorter than $8 \eta$), by means of Eq. (\ref{eq:epsilonFib}). The estimated value is represented by the green line in Fig. \ref{fig:epsilonfib}, with the error represented by the green shadow. This estimate is validated using PTV. Since PTV resolution does not allow a direct estimation of $\epsilon$ at the viscous scales, we estimate $\epsilon$ indirectly using the formula $\epsilon = - S_3^\pparallel / \left( \frac{4}{5} \,  \, r \right)$ that holds for each separation within the inertial range, under the assumption of isotropic turbulence. The latter is approximately satisfied for the flow we generate. The symbols in Fig.~\ref{fig:epsilonfib} represent the PTV estimates showing that the FTV value matches the PTV estimate for $r < L$: this proves that a small rigid fiber is able to estimate $\epsilon$.

\begin{figure}[t]
\includegraphics[width = \columnwidth]{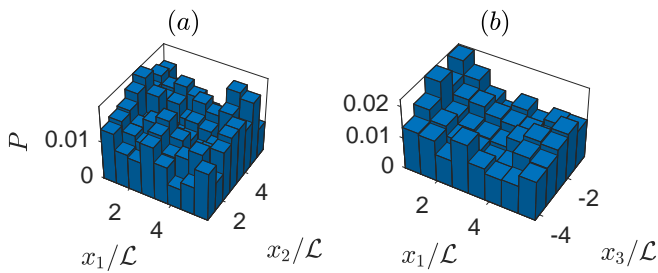}
\caption{\label{fig:fibBinningPos} Probability of the fiber positions from one of the experiments. The $x_1-x_2$ plane (a) and $x_1-x_3$ plane (b) are respectively orthogonal and parallel to the main camera direction. The fiber length is $c/\mathcal{L} = 0.76$ and the turbulence is generated by rough wheels rotating at $400\,\mathrm{rpm}$.}
\end{figure}

\begin{figure}[t]
\includegraphics[width = \columnwidth]{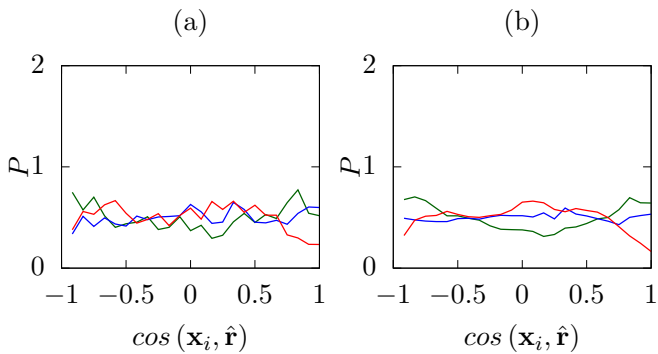}
\caption{\label{fig:fibBinningOri} PDF of the cosine of the angle between the lab coordinates $\textbf{x}_i$ and the fiber orientation unit-vector $\hat{\textbf{r}} = \textbf{r}/|\textbf{r}|$ for the whole observation volume (a) and conditioned on the region in the vicinity of the impellers ($x/\mathcal{L} < 0.71$) (b). The axis $\textbf{x}_3$ points toward the main camera direction, while $\textbf{x}_1$ and $\textbf{x}_2$ are orthogonal to it. The fiber length is $c/\mathcal{L} = 0.76$ and the turbulence is generated by rough wheels rotating at $400\,\mathrm{rpm}$.}
\end{figure}

\paragraph*{Random sampling and orientation.} In order to rule out a possible bias of the statistics due inhomogeneous sampling, the fibers must visit the spatial domain uniformly and stay randomly oriented. To check that the fibers sample the whole domain homogeneously, we analyze the normalized bin-counting histogram of the fiber position, that is an empirical estimator for the probability to observe a fiber at a given location. Fig. \ref{fig:fibBinningPos} shows that the fibers visited the whole domain in a homogeneous way, both considering the plain orthogonal (a) and parallel (b) to the main camera direction. By analyzing the probability of the cosine of the angle between the laboratory coordinates ($\textbf{x}_i$) and the fiber orientation vector ($\hat{\textbf{r}} = \textbf{r}/|\textbf{r}|$), we demonstrate (fig. \ref{fig:fibBinningOri} (a)) that the fibers are on average randomly oriented. Fig. \ref{fig:fibBinningOri} (b) shows the probability obtained by conditioning the analysis to one of the edges of the observation volume, where the flow is significantly inhomogeneous due to the vicinity of the impellers. No significant differences between the orientation probability measured in the whole domain and close to the impellers are appreciated, implying that the fibers sample the whole range of directions. There is some scatter due to limited statistics of a single experiment and when conditioning on a sub-volume of the domain. Other experiments (not shown) display similar scatter and no preferential orientation.

\begin{figure}[t]
\includegraphics[width = \columnwidth]{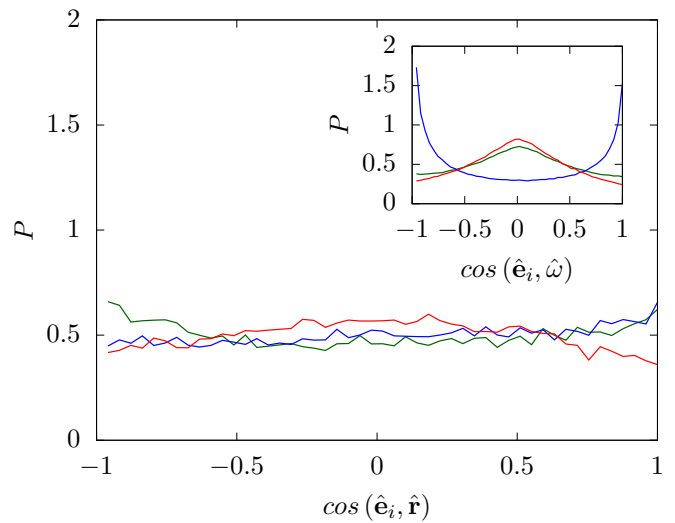}
\caption{\label{fig:fibAlign} PDF of the cosine of the angle between the strain rate tensor $\textbf{S}$ eigenvectors $\hat{\textbf{e}}_i$, and the fiber orientation unit-vector $\hat{\textbf{r}} = \textbf{r}/|\textbf{r}|$; the eigenvectors $\hat{\textbf{e}}_i$ are associated to the eigenvalue $\lambda_i$, such that $\lambda_1 \geq \lambda_2 \geq \lambda_3$ (green, blue and red lines respectively); the inset shows the PDF of the cosine of the angle between the strain rate tensor eigenvectors and the vorticity direction. The fiber length is $c/\mathcal{L} = 0.57$, while the scale $\Delta$ at which \textbf{S} is measured is $\Delta/L = 0.86$. The turbulence is generated by rough wheels rotating at $400\,\mathrm{rpm}$.}
\end{figure}

\paragraph*{Alignment between the fiber and the flow.} Random orientation and distribution in the observation volume does not imply that the fibers do not preferentially align with the local flow. We investigate this aspect by linking the local flow field to the fiber orientation. To this end, we perform an experiment in which fibers whose length falls into the inertial range of turbulence and tracer particles are tracked simultaneously (see Appendix \ref{sec:Fiber tracking technique} for a detailed discussion on the tracking technique). We measure the strain rate principal directions in a Lagrangian frame of reference attached to the fiber center at a scale larger than the fiber length. The spatial velocity derivatives along trajectories are evaluated following the approach proposed by \citet{luthi2005lagrangian} for particle tracers, that consists in considering all the tracers inside a sphere of given diameter centered at the fiber center of mass. Fig. \ref{fig:fibAlign} shows  negligible alignment with the principal strain directions for inertial range fibers. The inset shows the well-known vorticity-strain preferential alignment of the velocity field \citep{ashurst1987alignment,luthi2007lagrangian} which confirms the reliability of our coarse-grained strain-rate tensor evaluation. Some evidence of alignment between either rigid \citep{pujara_voth_variano_2019} or flexible \citep{picardo2020dynamics} fibers with length within the inertial range with the local flow is documented, and is thought to be at the origin of the spinning of long rigid fibers \citep{oehmke2021spinning}. \citet{pujara_voth_variano_2019} and \citet{picardo2020dynamics} numerically tracked infinitely slender passive fibers in an undisturbed turbulent flow field, hence neglecting the full dynamics arising from the two-way coupled interaction between the fiber and the flow. Conversely, in our experiments and simulations, the fiber actively modifies the surrounding flow, possibly explaining some of the differences. Moreover, as explained above, since the local flow is modified by the presence of the fiber, in our work the strain rate tensor is coarse-grained at a scale somewhat larger than the fiber length. This may lead to a weakening of a possible alignment between the fiber and the flow. Notwithstanding, Fig.\,\ref{fig:fibAlign} displays a slight tendency of the fiber to align with $\hat{\textbf{e}}_1$ and anti-align with $\hat{\textbf{e}}_3$, which is consistent with the one documented in \citep{pujara_voth_variano_2019,picardo2020dynamics}. With regards to short fibers (shorter than $8\eta$) we know from the experiments by \citet{ni2015measurements} that preferential alignment between the fiber orientation and the main strain direction $\hat{\textbf{e}}_1$ occurs. However, we have shown above by DNSs and experiments that this alignment does not significantly bias the measurement of the observable under consideration, namely the average turbulent dissipation rate. We also note that \citet{ni2015measurements} showed that the average tumbling rate (proportional to $\epsilon$) is underestimated by at most $20\%$ when conditioning the statistics only to fiber perfectly aligned to the main strain direction. This indicates that an accurate measurements of the average turbulent dissipation rate is still feasible despite some preferential alignment.

\section{\label{conclusion}Summary and perspectives}
We combined fully-resolved Direct Numerical Simulations and accurate laboratory experiments to show how rigid fibers can be used to measure two-point statistics of turbulence. This conclusion holds true for both spatial and temporal observables. For a fiber to be a proxy of turbulence eddies, two simple conditions must be fulfilled: i) the fiber length has to be comparable to the size of the eddy under consideration (i.e. the fiber length has to belong to the inertial range of scales); ii) the fiber inertia has to be negligible. Once both conditions are satisfied, the fiber velocity difference evaluated at the two fiber ends, projected along a transverse-to-the-fiber direction, is statistically equivalent to the unperturbed (i.e. evaluated in the absence of the fiber) flow transverse velocity difference computed at the scale of the fiber.

To be more specific, the numerical part of this study reveals how the Probability Density Function (PDF) of the fiber-based transverse velocity excursions between points separated in space by a distance within the inertial range of scales matches the same observable  obtained via standard Eulerian measurements of the unperturbed flow field. We found a similar agreement  by comparing the second- and third-order structure functions of the unperturbed flow velocity field with the same observables built by measuring the fiber end velocities. 

It is worth emphasizing that the above conclusions are not a trivial consequence of the no-slip condition imposed on the fiber. The no-slip condition indeed trivially imposes that all points of the fiber move at the same velocity of  the flow. However, because the fiber does not evolve as a tracer (owing to the fiber inextensibility condition and, in general, to its finite inertia) the flow velocity is locally modified by the presence of the fiber. The nontrivial finding here is that within the assumptions above, the back reaction of the fiber to the flow along a transverse direction is negligible, so that one can consider the transverse fiber-based velocities equal to the local transverse flow velocity.

Fibers also allow one to access the temporal properties of turbulence eddies belonging to the inertial range of scales. In this respect we showed that the eddy-turnover time at a given spatial scale, say $r$, can be easily measured by analyzing the tumbling time of a fiber of length $r$, measured along the Lagrangian trajectories. The two times match provided that the fiber inertia is sufficiently small. 

Considering sufficiently short fibers belonging to the viscous range, we showed through DNS that transverse fiber-based velocity increments can be used to build transverse fiber-based velocity derivatives, if the two following conditions apply: i) the fiber inertia is sufficiently small, and ii) the fiber length is shorter than  $\sim \, 8 \eta$. Furthermore, we showed that these quantities are what one needs to measure the flow energy dissipation rate in terms of our fiber-based measurements.

Our numerical study thus suggested the possibility of tracking rigid fibers as a convenient alternative to measure two-point statistical properties of turbulence in the inertial range of scales as well as flow field derivatives along the transverse direction of the fibers in the viscous sub-range. We took up the challenge and carried out laboratory experiments with custom-made polymeric fibers in a turbulent flow. The results fully confirm the scenario from the numerical analysis and also give birth to a new technique termed Fiber Tracking Velocimetry (FTV) capable to easily and economically access two-point statistics of turbulence.

The latter possibility appears of paramount importance in experimental turbulence because of the different reasons we discuss below. The advantages of using fibers instead of particles as Lagrangian tracers in measuring turbulent flows depend on the particular measurement one wants to undertake. In the laboratory environment, the measurement of large scale turbulence statistics by using tracer particles is feasible and convenient: indeed, when considering an enclosed domain (such as a water tank) the tracer dispersion hardly prevents obtaining convergent statistics. However, when investigating small scale properties of the flow, the measurements get extremely problematic. In fact, to increase the flow field resolution by simple particle tracking one needs to increase the concentration of tracer particles. If the particles are uniformly distributed in a fixed observation volume, the average separation between particle couples decreases as the inverse of the cube root of the particle concentration. In other words, to increase $10$ times the flow field resolution, $1000$ times more particles are needed, resulting in a technological insurmountable obstacle. In conventional particle tracking experiments, a limit of $\sim 10^3-10^4$ tracer particles simultaneously present in the field of view is imposed mainly by ambiguities of particle positions arising from particle image overlap on the image chip of the camera. Ambiguities can be tolerated to a certain extent; they become, however, prohibitive with further increase of the seeding density. To evaluate dissipation, classical approaches rely on the local velocity gradient tensor that is estimated using several particles within a Kolmorogorov size volume. The higher the turbulence intensity, the greater is the flow field resolution needed to investigate small scale flow structures resulting in a bottle neck when measuring high Reynolds number flows.

The novel approach proposed here allows to measure spatial derivatives along the fiber trajectory looking also at its orientation other than at the position. In principle, this can be done through FTV without increasing the tracer concentration, in extremis using only one fiber. Moreover, recent numerical findings \cite{cavaiola_olivieri_mazzino_2020} demonstrate the possibility of measuring the instantaneous velocity gradient tensor by means of suitable assemblies of rigid fibers, representing a potential tipping point in the field of experimental turbulence.

By measuring the two-point statistics within the inertial range, we prove the FTV reliability to measure finite scale velocity differences. In this regard, fibers of different lengths can be considered as a proxy of the celebrated Richardson cascade: when having sufficiently small inertia, the fiber is captured by the \textit{whirls}, becoming a Lagrangian tracer that moves as the eddies of its own size. 

In a laboratory environment, finite-size fibers may not always be more handy to measure inertial range scaling laws than traditional tracer particles. However, the advantages related to FTV become fully relevant when considering field measurements in unbounded domains, such as the ocean and the atmosphere. There, the natural tendency of tracers to increase their mutual distance (by virtue of Richardson's law) precludes easy measures of small-scale two-point turbulence statistics. The brute-force approach, consisting in increasing the number of available tracers to ensure a given separation to be always covered by the Lagrangian points, is not realizable in practice. In the ocean, a tracer is indeed a buoy which is costly, thus preventing a massive use. We thus expect our method to find broad application and to provide improved statistics of turbulence in environmental flows that are of paramount importance for weather and climate predictions.

\section{Acknowledgments}
The authors acknowledge computer time provided by the Scientific Computing section of Research Support Division at OIST.

\appendix

\section{Numerical simulations}
\label{app:DNS}
Here we provide additional information on the numerical simulations whose results are presented in Sec.~\ref{sec:DNS}.

Direct numerical simulations of sustained homogeneous isotropic turbulence in a three-periodic domain are performed with rigid fibers suspended within. 
The fluid flow is governed by the incompressible momentum and continuity equations:
\begin{equation}
\frac{\partial \textbf{u}}{\partial t} + \textbf{u} \cdot \nabla \textbf{u} = -\frac{1}{\rho_\mathrm{f}} \nabla p + \nu \Delta \textbf{u} + \fb_\mathrm{FOR} +  \fb_\mathrm{FIB},
\end{equation}
\begin{equation}
\nabla \cdot \textbf{u} = 0,
\end{equation}
where $\textbf{u}\left( \textbf{x},t \right)$ and $p\left( \textbf{x},t \right)$ are the fluid velocity and pressure fields, $\rho_\mathrm{f}$ and $\nu$ are the fluid volumetric density and kinematic viscosity, and $\fb_\mathrm{FOR}$ and $\fb_\mathrm{FIB}$ are forcing terms used to sustain turbulence and model the presence of the suspended fibers~\cite{rosti2018flexible,rosti2019flowing}.
In particular, the turbulent flow is sustained using the spectral forcing scheme by~\citet{eswaran1988forcing}, where energy is injected randomly at low wavenumbers (in our case, within a spherical shell with radius $k=2$) by means of a Ornstein-Uhlenbeck process.

The fiber dynamics is governed by the Euler-Bernoulli beam equation and by the inextensibility constraint
\begin{equation}
\Delta \widetilde{\rho} \, \frac{\partial^2 \textbf{X}}{\partial t^2} = \frac{\partial }{\partial s} \left( T \frac{\partial \textbf{X}}{\partial s} \right) - \gamma \frac{\partial^4 \textbf{X}}{\partial s^4} - \textbf{F},
\end{equation}
\begin{equation}
\frac{\partial \textbf{X}}{\partial s} \cdot \frac{\partial \textbf{X}}{\partial s} = 1,
\end{equation}
where $\textbf{X} \left( s, t \right)$ is the position of the fiber point as a function of the curvilinear coordinate $s$ and time $t$. In the previous equations,  $\Delta \widetilde{\rho} = \widetilde{\rho}_\mathrm{s} - \widetilde{\rho}_\mathrm{f}$ is the difference between the linear density of the fiber and the fluid one, $T$ is the tension needed to enforce the fiber inextensibility and $\gamma$ is the fiber bending rigidity (for a homogeneous fiber, it is the product of the elastic modulus and the second moment of the area). When $\gamma$ is chosen to be very large, the fiber behaves effectively as a rigid one, i.e., with negligible deviation of its end-to-end distance from the nominal fiber length. Finally, $\Fb$ is the fluid-structure coupling term. Since we consider dispersed fibers, freely-moving boundary conditions are imposed at both ends, i.e., $ \partial_{ss} \Xb |_{s=0,c} = \partial_{sss} \Xb |_{s=0,c} = T |_{s=0,c} = 0$.

The fluid and fiber are coupled by an immersed boundary method (IBM) where the no-slip condition $\dot{\Xb} = \ub \left( \textbf{X} \left( s,t \right),t \right)$ is enforced by means of a singular force distribution~\cite{peskin2002}.
In particular, we use the method originally proposed by~\citet{huang2007simulation} and later modified by~\citet{banaei2019numerical}.
At each timestep, the fluid velocity at the position of the fiber point, $\textbf{U} = \ub \left( \textbf{X} \left( s,t \right),t \right)$, is found by interpolating the values of $\ub$ at the nodes of the Eulerian grid surrounding the Lagrangian point
\begin{equation}
\textbf{U} \left( \textbf{X} \left( s,t \right),t \right) = \int \textbf{u} \left( \textbf{x},t \right) \delta \left( \textbf{x} - \textbf{X}\left( s,t \right) \right) \mathrm{d}^3 \textbf{x},
\end{equation}
where the $\delta$ is the Dirac delta function.
The interpolated velocity $\Ub$ is used to compute the fluid-structure forcing
\begin{equation}
    \Fb(s,t) = \kappa \, (\Ub - \dot{\Xb}),
\end{equation}
where $\kappa$ is a large negative constant~\cite{huang2007simulation}. Finally, $\Fb$ is transferred to the fluid as 
\begin{equation}
\fb_\mathrm{FIB} \left( \textbf{x},t \right) = \frac{1}{\widetilde{\rho}_\mathrm{f}} \int_s \textbf{F} \left( s,t \right) \delta \left( \textbf{x} - \textbf{X}\left( s,t \right) \right) \mathrm{d} s.
\end{equation}
Both the interpolation and spreading operations feature the Dirac operator, which in our numerical simulations is transposed into the regularized $\delta$ function proposed by~\citet{roma1999}.

The problem is solved numerically using the fractional step method on a staggered grid with the second-order finite-difference scheme in space and the third-order Runge-Kutta scheme in time~\cite{kim2007penalty}. Additionally, the Poisson equation enforcing the incompressibility constraint is solved using the Fast Fourier Transform. Specifically, we employ the same numerical procedure already used for moving and deforming filaments in laminar or turbulent flows in Refs.~\cite{rosti_brandt_2017a,rosti2018flexible,rosti2019flowing,banaei2019numerical,cavaiola_olivieri_mazzino_2020}, to which the reader is referred to the further information. 

Simulations are performed using a Cartesian uniform mesh in a rectangular tri-periodic box of size $L = 2 \pi$, with $128$ grid points per side. The grid size is sufficient to resolve an inertial range of scales clearly showing the $4/5$th Kolmogorov law. Doubling the resolution in all directions leads to a negligible  change of the results.

\section{\label{sec:Fiber fabrication}Fiber fabrication}
Hereafter, we describe the two methodologies we employed to fabricate the polymeric fibers. Both are conceived to be low-cost, and do not need specific lab equipment. The first method consists in producing fibers tagged with a fluorescent dye at their edges, allowing to track them by applying optical filters to the cameras. The second method consists in using fully dyed fibers, whose edges can be tracked resorting to a customize image segmentation procedure. In both cases, the fibers are characterized by a high rigidity and strength in comparison to the flow forcing. The main limitation of the first methodology is that only slender fibers longer than $5\,\mathrm{mm}$ can be easily fabricated using low budget lab equipment, while the second methodology allows to  hand-craft nylon fibers that can be even shorter than $2\,\mathrm{mm}$. Notwithstanding, the first methodology allows a greater accuracy in the fiber tracking procedure.

\paragraph*{Fibers with dyed edges.}
We start by detailing the procedure developed to produce rigid fibers tagged with fluorescent dye at their ends, suitable for tracking their edges and measure transverse velocity differences between them. The chosen material is polydimethylsiloxane (PDMS), whose physical properties are listed in Table~\ref{tab:material_prop}, while for the fluorescent dye we use rhodamine B. PDMS is suitable for our purposes since it is characterized by similar density as water, thus making the fibers almost neutrally buoyant. Moreover, PDMS has a refractive index similar to water, hence the fiber body is almost transparent when immersed in the fluid while the edges are still visible (Fig~\ref{fig:parVSfib}). In addition, the diffusion process of rhodamine B in the PDMS matrix is slower than the polymerization, meaning that it is possible to dye only a certain part of a PDMS sample without deeply penetrating into the matrix. The standard procedure to produce PDMS samples consists of mixing an elastomer with a curing agent (ratio 10/1) by stirring the solution for at least 10 minutes; the preparation is then desiccated in a vacuum chamber and placed in a Petri dish. If any bubbles form while the solution is poured, a pipette can be used to burst them out. Finally, the mixture is baked in an oven at $\SI{80}{\celsius}$ for at least 3 hours. In our case, to produce fibers with the ends tagged by the fluorescent dye, we modify the standard protocol and perform the following procedure:
\begin{enumerate}
\item two different beakers of PDMS samples are brought to the liquid state, one of pure PDMS and the other mixed with rhodamine B;
\item a first thin layer of rhodamine-dyed PDMS is poured in a Petri dish and baked in the oven for 15 minutes, to obtain a solution not completely cured;
\item a second thicker layer of pure PDMS is added on top of the first, and the overall sample is cooked for additional 2 hours; at this point of the procedure, the first layer is completely cured;
\item a third rhodamine-dyed PDMS layer is added and the whole sample is cooked for at least 1 hour to ensure its complete polymerization;
\item the sample is pierced through the layers using a special cylindrical puncher to pull out the fibers.
\end{enumerate}
Note that, the baking times are shortened and optimized to prevent the PDMS solution from completely curing before the addition of the following layer. With this modification, the above layer can stick to the lower one without allowing the dye to diffuse in the pure PDMS layer. By following this protocol, three-layered fibers are obtained. Overall, from a single sample of about $10 \times 10 \, \mathrm{cm}^2$, it is possible to punch more than $200$ fibers. The fiber length $c$ can be controlled by changing the thickness of the middle layer. In this work, $c$ will vary from $6$  to $11\,\mathrm{mm}$. The diameter can be controlled by changing the puncher size. In our case, we produce fibers with diameter $d = 1\,\mathrm{mm}$, which is sufficiently large to ensure an essentially rigid behavior and a sufficient slenderness $c/d \gg 1$.

\paragraph*{Fully dyed fibers.}
Fully dyed fibers are fabricated by cutting a white reflective nylon string. While the fiber diameter is fixed by the nylon string thickness, the fibers length can be controlled in the cutting phase. The physical and mechanical properties of Nylon are also listed in Table~\ref{tab:material_prop}. This material was chosen because of its availability; moreover, its extremely high Young modulus ensures the fibers rigidity even with small diameters. Since Nylon mass density differs significantly from water, a suitable amount of salt has to be dissolved in water to achieve the neutral buoyancy. In addition, Nylon has a refractive index that is strongly different from water: consequently Nylon fibers are well visible when illuminated by the laser light.

\begin{table}[h]
\caption{\label{tab:material_prop} Physical properties of the employed material. The table reports the volumetric density $\rho_\mathrm{s}$, the Young modulus $E$, the Poisson ratio $\nu_\mathrm{P}$, the ultimate tensile stress $\sigma_\mathrm{u}$ and the refractive index $\mathit{IR}$.}
\begin{ruledtabular}
\begin{tabular}{@{}cccccc@{}}
& $\rho_\mathrm{s}$ & $E$ & $\nu_\mathrm{P}$ & $\sigma_\mathrm{u}$ & $\mathit{IR}$\\
 & $\mathrm{kg\,m^{-3}}$ & $\mathrm{MPa}$ & $-$ & $\mathrm{MPa}$ & $-$\\
\colrule
PDMS & $965$ & $0.360 \div 0.870$ & $0.5$ & $2.24$& $1.4$\\
nylon & $1140$ & $200 \div 400$ & $0.41$ & $82$ & $1.53$\\
\end{tabular}
\end{ruledtabular}
\end{table}

\section{\label{sec:Fiber tracking technique}Fiber tracking technique}
The position and orientation of each fiber is determined assuming that they are rigid and almost one-dimensional bodies: consequently, only two distinct points are needed to determine its position and orientation. In order to discern unequivocally two points of the fiber, its edges are considered, and their coordinates determined using the OpenPTV algorithm. When multiple fibers are present in the same image sequence, a proper postprocessing routine (programmed in MATLAB) is used to distinguish each single fiber.

To determine the edge coordinates in the image space, two different methodologies have been employed: the first one is suitable to track transparent fibers tagged at the edges with a fluorescent dye, while the second allows to define the edges position when using fully dyed fibers.

\begin{figure}[h]
\centering
\subfigure[{ }\label{fig:parVSfib_1}]
{\includegraphics[width = 2.7cm]{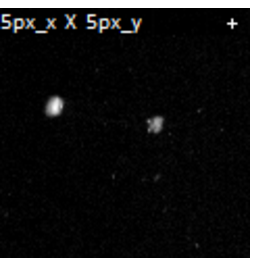}}
\,
\subfigure[{ }\label{fig:parVSfib_2}]{\includegraphics[width = 2.7cm]{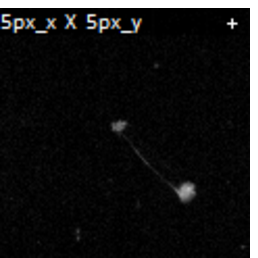}}
\,
\subfigure[{ }\label{fig:parVSfib_3}]{\includegraphics[width = 2.7cm]{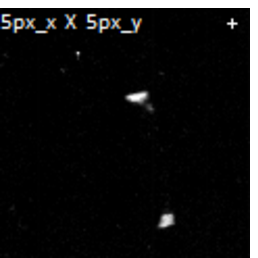}}
\caption{Details of three fibers as seen though optical filters: only the dyed edges are visible. A threshold of $5 \times 5$ pixels is imposed on the pixel clusters size to ensure not to detect dust and small residual particles.}
\label{fig:parVSfib}
\end{figure}

\paragraph*{Edges detection by optical filters.}
The first method employed to determine the edges coordinates in the image space, is inspired by \citet{klein2012simultaneous}, \citet{bellani2012shape} and \citet{bordoloi2017rotational}, who immersed tracer particles inside hydro-gel spheres and cylinders, to determine their angular velocity performing PIV analysis on the embedded tracers. Similarly, we tagged the edges of the fibers with a fluorescent dye, and performed standard PTV to track their position. By shielding the cameras with optical filters that remove the laser light frequency, only the fluorescent edges remain visible. Three snapshots of a camera image are shown in Fig.~\ref{fig:parVSfib}: some impurities (probably stemming from small fluorescent tracer particles from previous experiments) are still visible. Since the edges are considered as particle clusters by the OpenPTV software, a threshold on the pixel cluster size may be adopted to ensure the correct detection of the fiber edges only. By adopting a suitable postprocessing algorithm (see \cite{luthi2005lagrangian}) the Lagrangian velocities at the two fiber ends are evaluated. Eventually, only the couples of points located at a distance that is the fibers length (known a priori) are considered. Mismatches are still possible when couples of edges belonging to two distinct fibers are separated by the fiber length: notwithstanding, this case is unlikely because the maximum local concentration is extremely low; hence, this potential error does not affect the statistics significantly.

\begin{figure}[h]
\centering
\subfigure[{original}\label{fig:fib_org}]
{\includegraphics[width = 4cm]{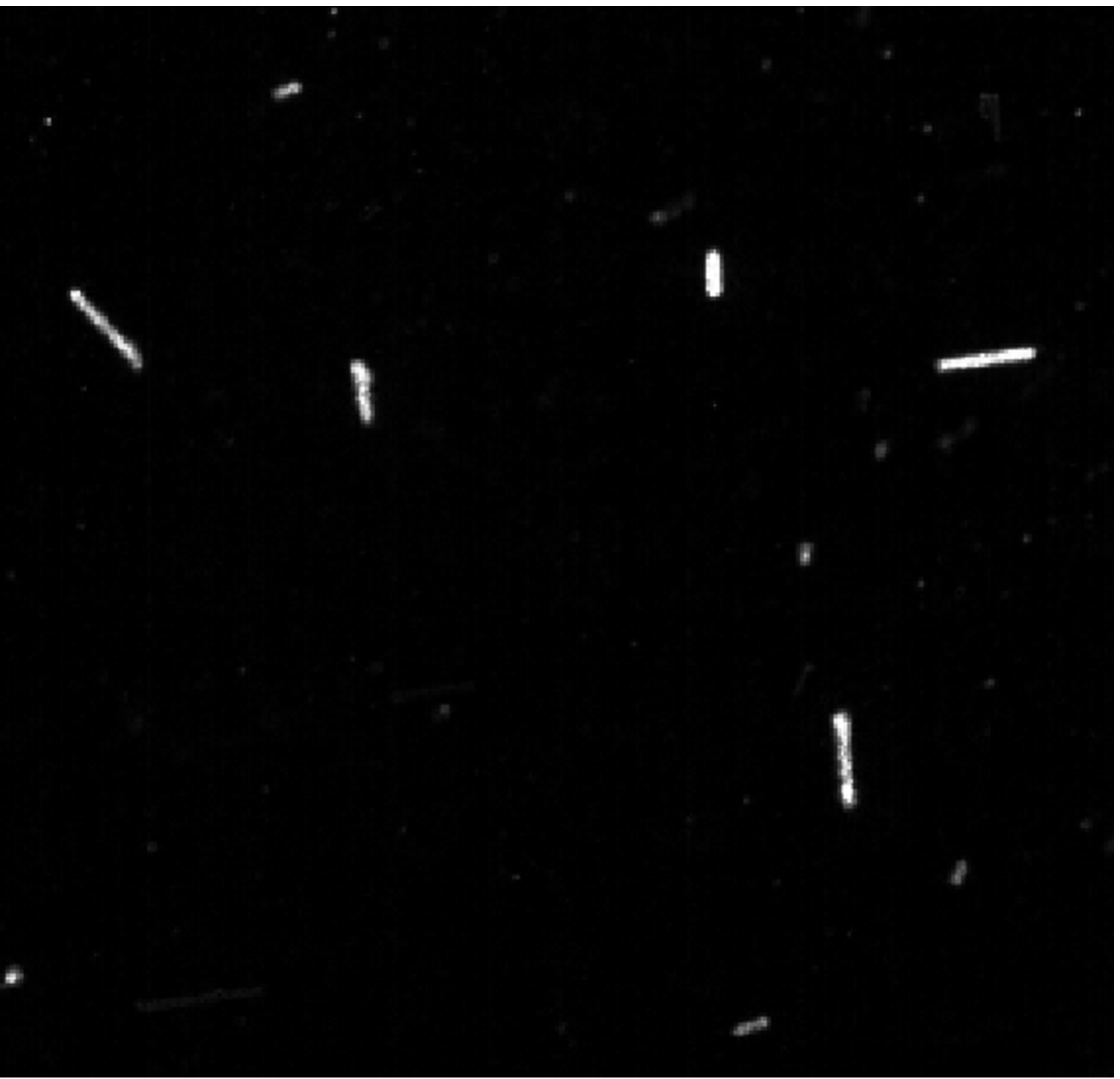}}
\,
\subfigure[{segmented}\label{fig:splitter}]{\includegraphics[width = 4cm]{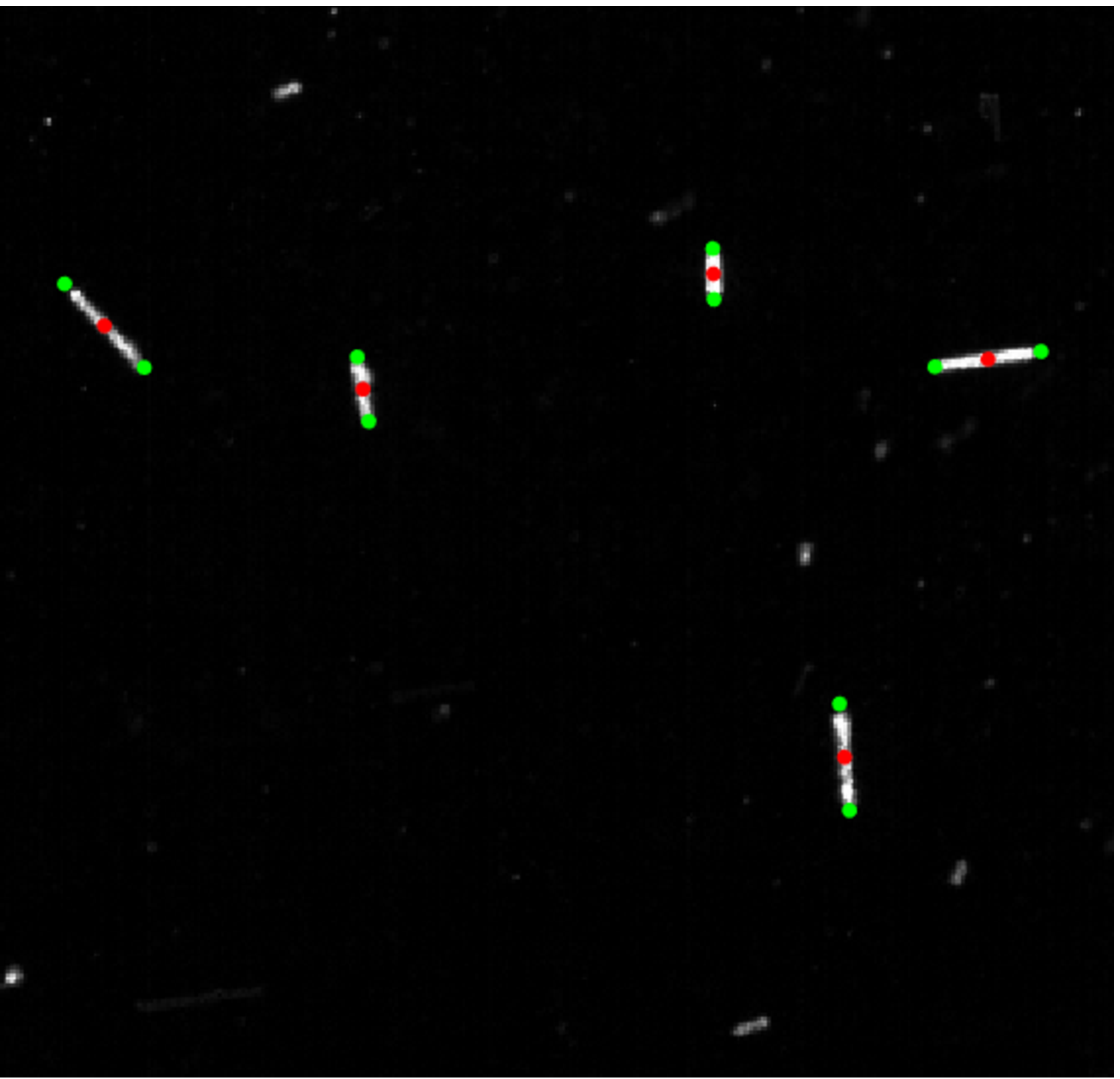}}
\caption{Detail of an image in which fully dyed fibers are visible; the red dots of image b are the fibers centroids while the green dots are the fibers edges.}
\label{fig:segmentation}
\end{figure}

\paragraph*{Edge detection by customized image segmentation.}
OpenPTV allows to employ a customized image segmentation when defining the particle position in the image space. Providing to the software the position of the fiber edges and centers in the image space, it is possible to perform detection and tracking considering them as particles. Fig.~\ref{fig:fib_org} shows a detail of an image of fully dyed fibers as it is recorded. The image segmentation process consists in the following steps: i) a time-mean filter is applied to subtract the (static) background light to the image sequence; ii) the contrast is stretched to evidence the fibers and the images are binarized to detect the pixel clusters; iii) clusters smaller than a threshold are discarded in order to remove eventual dust or small particles, while on pixel clusters representing the fibers an ellipse is fitted; iv) the major, the minor axes and the centroid are evaluated for each fiber: the centroid is the fiber geometric center, while the vertices of the ellipse represent the edges. From the segmented images, both the trajectories of the fibers centers and the edges can be determined independently with OpenPTV.

The two methodologies present advantages and drawbacks. The first method allows to track only the fiber edges; however, being based on optical principles, there is no need of tuning additional parameters. Conversely, in the second methodology one needs to adjust the segmentation parameters to the light condition, which changes significantly between different experiments and camera position and orientation. However, by the use of customized image segmentation, it is possible to track very small fibers (in our experiments, slightly shorter than $3\,\mathrm{mm}$). This allows to investigate the behavior of fibers close to the Kolomogorov lengthscale. Indeed, here it is not necessary to optically distinguish the two fibers edges, but only an elongated pixel cluster; moreover, the handcrafting of extremely short and slender PDMS fibers tagged with dyed edges is technologically challenging.

\paragraph*{Simultaneous fiber and tracer tracking.}
To analyze possible preferential alignment with the local flow, simultaneous fiber and tracer particle tracking is employed. To track fibers and particles simultaneously we adopted a set-up inspired by the one used by \cite{michalec2017zooplankton}. Four synchronized cameras are employed, three of which are used to track the particle tracers, while the fourth is equipped with an image splitter, namely an optical arrangement that allows stereoscopic imaging using one single camera (see Fig. \ref{fig:splitter}). The fourth camera is shielded with an optical filter, so that, as in the previous experiment, only the fluorescent fiber edges are visible.

\begin{figure}[h]
\includegraphics[width = \columnwidth]{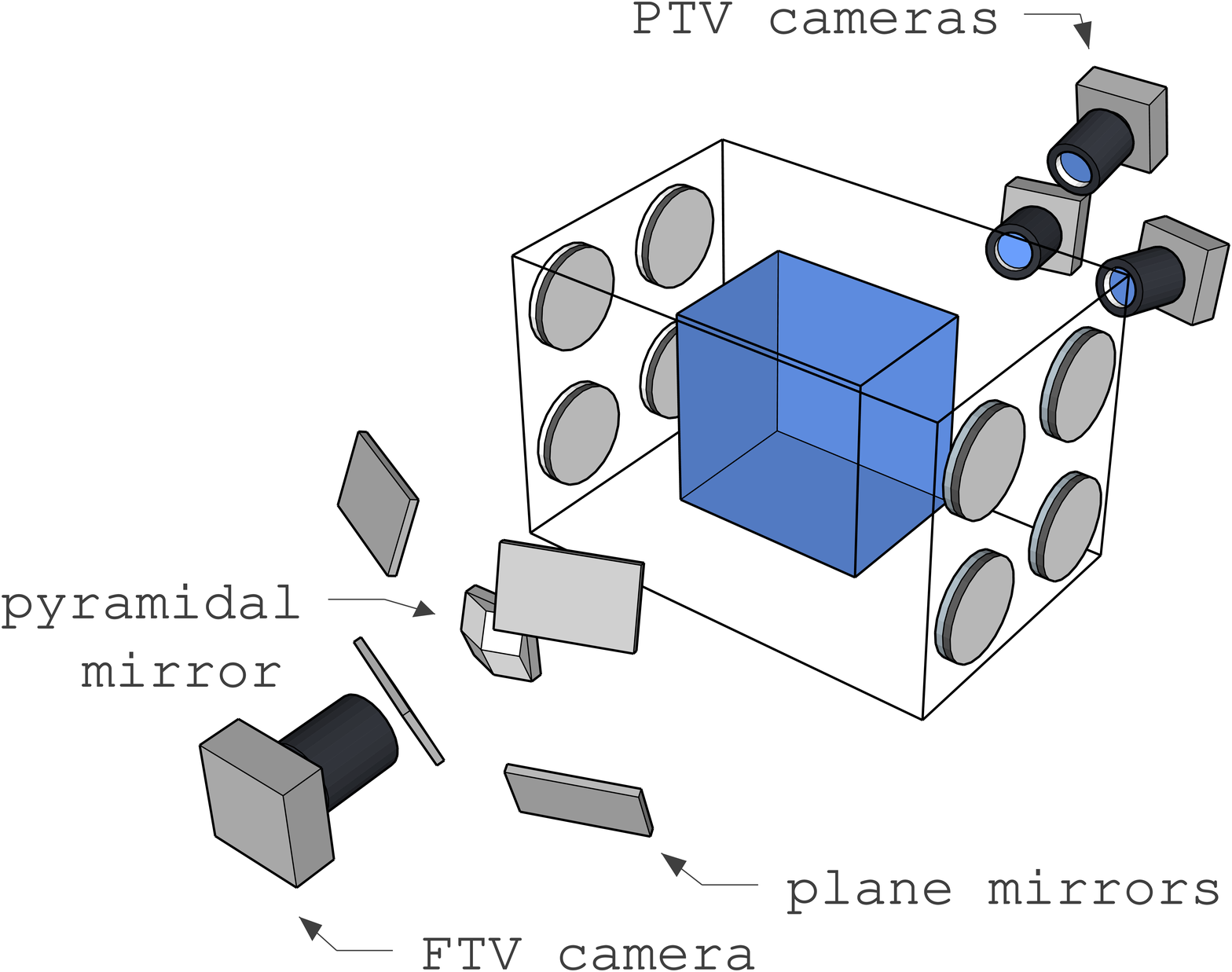}
\caption{\label{fig:splitter} Schematic representation of the simultaneous fiber-particle tracking system. Four synchronized cameras are employed, three of which for particle tracking velocimetry, while the fourth, equipped with an image splitter and a green-cutting off filter, records the motion of the fibers. The image splitter consists of of four plate mirrors that project the reflection of the observation volume from different view-points on a pyramidal mirror and subsequently onto the camera sensor.}
\end{figure}

\newpage
% \bibliography{sample} 

\begin{thebibliography}{55}
\bibitem [{tur()}]{turbPropertiesNote}%
  \BibitemOpen
  \href@noop {} {}\bibinfo {note} {Rotating wheels at $400\,\mathrm{rpm}$
  produce a turbulence such that the root mean square velocity $u' =
  94.7\,mm/s$, the integral scale $\mathcal{L} = 15\,mm$ and the average
  turbulent dissipatio rate obtained by balancing the production $\epsilon =
  u'^3/\mathcal{L} = 0.057\,m^2 s^{-3}$. The Kolmogorov length and time
  respectively $\eta = (\nu^3/\epsilon)^{1/4} = 0.065\,mm$ and $\tau_\eta =
  (\nu/\epsilon)^{1/2} = 0.0042\,s$, while the Taylor microscale $\lambda =
  \left( 15\,\nu/\epsilon\right)^{1/2}\,u' = 1.5\,mm$. The Reynolds number
  based on the Taylor microscale $Re_{\lambda} = 146$, while the one based on
  the integral length scale $Re_{\mathcal{L}} = 1410$.}\BibitemShut {Stop}%
\end{thebibliography}


\end{document}